\documentclass[a4paper]{jpconf}

\begin{document}

\title{Thermal neutrinos from pre-supernova}

\author{A. Odrzywolek$^1$, M. Misiaszek$^1$, M. Kutschera$^{1,2}$}

\address{$^1$ M. Smoluchowski Institute of Physics, Jagiellonian University, Reymonta 4, Cracov, Poland}
\address{$^2$ H. Niewodniczanski Institute of Nuclear Physics, 
PAS, Radzikowskiego 152, Cracov, Poland}

\ead{odrzywolek@th.if.uj.edu.pl}

\begin{abstract}
We would like to discuss prospects for neutrino observations
of the core-collapse supernova progenitor during neutrino-cooled stage.
We will present new theoretical results on thermal neutrino and antineutrino spectra
produced deep inside the pre-supernova core. Three competing processes: pair-, photo and
plasma-neutrino production, are taken into account. The results will be used to
estimate signal in existing and future neutrino detectors. Chance for supernova prediction
is estimated, with possible aid to core-collapse neutrino and gravitational wave detectors
in the form of early warning.
\end{abstract}

In our short contribution we would like to answer
some comments and questions asked during poster session \cite{poster_AOdrzywolek}.
Two days of the silicon burning with $\langle L_\nu\rangle = 3 \times 10^{45}$ erg/s quoted in \cite{OMK}
was randomly chosen typical example model \cite{woosley_1978}.
Duration of the Si burning depends on evolutionary track
of the massive star, particularly mass of the Si core at the onset of the
ignition. Total average neutrino luminosity $\langle L_\nu \rangle$ depends on (1)  the burning time $\tau_{\mathrm{Si}}$
and (2) amount of fuel (Si core mass $M_{\mathrm{Si}}\simeq M_{\mathrm{Fe}}$
):
\begin{equation}
\langle L_\nu \rangle = M_{\mathrm{Fe}}\; \Delta E_b  / \tau_{\mathrm{Si}}.
\end{equation}
$\Delta E_b$ is the nuclear binding energy difference
between fuel ("Si") and ash ("Fe"). From evolutionary calculations of \cite{woosley_rmp}
we get core mass in the range 1.2 \ldots 1.65 M$_\odot$ and $\tau_{\mathrm{Si}}$
from 18 days to 17 hours, 
respectively. 
Therefore $\langle L_\nu \rangle$
vary from $3.1 \times 10^{44}$  to $9.8 \times 10^{45}$ erg/s.

Electron antineutrinos from pair-annihilation are of particular interest \cite{gadzooks}.
One can quickly estimate fraction of given flavor from the following:
(1) reaction rate is proportional to squared matrix element 
$M^2 \sim 8 G_F^2 (C_V^2 + C_A^2)$, where mass term was dropped
(2) number of emitted neutrinos and antineutrinos is equal. 
Therefore, rate of the reaction rates is $R_e/R_{\mu,\tau} \simeq
({C^e_V}^2 + {C^e_A}^2)/({C^{\mu,\tau}_V}^2 + {C^{\mu,\tau}_A}^2) = 4.5$.
Total neutrino flux is: $F_{tot} = 2 R_e + 4 R_{\mu,\tau} = (2+4/4.5) R_e = 2.9 R_e$. Finally $F_{\bar{\nu}_e}/F_{tot} \simeq 0.35 \sim 1/3$, i.e. about one third
is emitted as electron antineutrinos. This statement may be 
altered by the neutrino oscillations.

\ack 
Supported by grant of Polish Ministry of Science and Higher Education No. 1~P03D~005~28.

\end{document}